\documentclass{emulateapj}

\shorttitle{Solar constraints on asymmetric dark matter}
\shortauthors{Lopes \& Silk}

\begin{document}


\title{Solar constraints on asymmetric dark matter}


\author{Il\'\i dio Lopes\altaffilmark{1,2,6}, Joseph Silk\altaffilmark{3,4,5,7}}

\altaffiltext{1}{Centro Multidisciplinar de Astrof\'{\i}sica, Instituto Superior T\'ecnico, 
Universidade Tecnica de Lisboa , Av. Rovisco Pais, 1049-001 Lisboa, Portugal} 
\altaffiltext{2}{Departamento de F\'\i sica,Escola de Ciencia e Tecnologia, 
Universidade de \'Evora, Col\'egio Luis Ant\'onio Verney, 7002-554 \'Evora - Portugal} 
\altaffiltext{3}{Institut d'Astrophysique de Paris, France} 
\altaffiltext{4}{Department of Physics, University of Oxford, United Kingdom} 
\altaffiltext{5}{Department of Physics and Astronomy, The Johns Hopkins University, Baltimore, MD21218}
\altaffiltext{6}{E-mail: ilidio.lopes@ist.utl.pt} 
\altaffiltext{7}{E-mail: silk@astro.ox.ac.uk}  


\begin{abstract}
The dark matter content of the Universe is likely to be a mixture of matter and antimatter, 
perhaps comparable to the measured asymmetric mixture of baryons and antibaryons.  
 During the early stages of the Universe, the dark matter particles are produced in a process similar to baryogenesis, and
dark matter freeze-out depends on the dark matter asymmetry and the annihilation cross section (s-wave and p-wave annihilation channels) of particles and antiparticles. In these $\eta-$parametrised asymmetric dark matter models ($\eta$ADM), the dark matter particles have an annihilation cross section close to the weak interaction cross section, and a value of dark matter asymmetry  $\eta$ close to the baryon asymmetry $\eta_B$. 
Furthermore, we assume that dark matter scattering of baryons, namely, the spin-independent scattering cross section, is of the same order as the range of values  suggested by several theoretical particle physics models used to explain the current unexplained events reported in the DAMA/LIBRA, CoGeNT and CRESST experiments. 
Here, we constrain $\eta-$parametrised asymmetric dark matter by investigating the impact of such a type of dark matter on the evolution of the Sun, namely, the flux of solar neutrinos and helioseismology. 
We find that dark matter particles with a mass smaller than 15 GeV, a spin-independent scattering cross section on baryons of the order of a picobarn,
and an $\eta-$asymmetry with a value in the interval $10^{-12}-10^{-10}$, 
would induce a change in solar neutrino fluxes in disagreement with current neutrino flux measurements. 
This result is also confirmed by helioseismology data. 
A natural consequence of this model is suppressed annihilation, thereby reducing the tension between indirect and direct  dark matter detection experiments, but the model also allows a greatly enhanced annihilation cross section.
All the cosmological $\eta-$asymmetric dark matter scenarios that we discuss have a relic dark matter density 
$\Omega h^ 2$ and baryon asymmetry $\eta_B$  in agreement with the current WMAP measured values,
$\Omega_{DM} h^2=0.1109\pm0.0056$  and  $\eta_B=0.88\times 10^{-10}$. 
\end{abstract}
\journalinfo{The Astrophysical Journal, 757, 130, doi:10.1088/0004-637X/757/2/130}


\maketitle

%
%
\section{Introduction\label{sec-intro}}
 
During the last three decades, overwhelming evidence has been found 
for  the existence of dark matter in the Universe. This achievement  is the result of a careful analysis 
of the available cosmological observational data, combined with a variety of well structured theoretical physical models 
coming from quite different and complementary research fields in  particle physics, cosmology and astrophysics. 
Such studies have led the way to  identification of the basic gravitational effects of dark matter, 
and their contribution to the formation of structure in the universe. 
In spite of this, the fundamental nature of the dark matter particles remains
a mystery. 

Current observational studies suggest that the matter present in the Universe
is composed  predominantly of  dark matter particles and baryons\citep{Munshi:2011jg}.
Recent measurements of the cosmic microwave background by the WMAP team \citep{Larson:2011ji} 
give   precise measurements of the dark matter density and the baryonic density:
$\Omega_{DM}h^2 =0.1109\pm0.0056$ and $\Omega_{B}h^2 =0.02258^{+0.00057}_{-0.00056}$.  
These observational studies also measure the imbalance between baryons and antibaryons,
i.e., the baryon asymmetry,  which is found to be equal to  $(0.88\pm 0.021)\times 10^{-10}$.
 
Among the most popular candidates for dark matter are
a group of particles that occur naturally in supersymmetric extensions of the standard model
of particle physics, usually called Weakly Interacting Massive Particles (WIMPs), 
the neutralino, the lightest supersymmetric stable particle being the typical example.
Significant constraints in the properties of WIMP candidates and similar particles have been made using 
stars\citep{2005PhR...405..279B}, such as the Sun
\citep[e.g.,][]{2010Sci...330..462L,2010ApJ...722L..95L,2010PhRvD..82h3509T,2010PhRvD..82j3503C},
sun-like stars \citep{2011MNRAS.410..535C,2011ApJ...733L..51C}
and neutron stars \citep[e.g.,][]{Kouvaris:2011tz,2011PhRvL.107i1301K,2010PhRvD..82f3531K,2008PhRvD..77d3515B}.
Although, WIMP candidates can be either dirac or majorana  particles,
usually, WIMPs are considered to be of the later type, as such these particles do not have an asymmetry 
such as the baryon asymmetry.
Recently, several authors \citep[e.g.][]{2006PhRvD..73k5003G,2009PhRvD..80c7702F,2005PhLB..605..228H} 
following previous work \citep[e.g.][]{Nussinov:1985ig,Kaplan:1992dn} have proposed a new type of matter, 
known as asymmetric dark matter,  which has an asymmetry identical to baryons \citep{2009PhRvD..79k5016K}.
Like WIMPs, such particles are non-relativistic massive particles that interact 
with baryons on the weak scale,  thereby having a  
sizeable scattering cross-section with baryons \citep{2009PhRvD..79k5016K},
but unlike WIMPs, such a type of dark matter is produced in the primordial Universe by a mechanism similar to baryogenesis.
This type of dark matter, much the same as baryons, is considered to have an asymmetry that we choose to represent by the parameter $\eta$. Hence, this dark matter is composed of an unqual amount of matter and antimatter \citep[e.g.][]{2006PhRvL..96d1302F}. Furthermore, these particles have a mass of the order of a few GeV \citep[e.g.][]{2011arXiv1102.5644K,2010PhRvD..82e6001C}. 
We will refer to this type of dark matter as $\eta-$parametrised asymmetric dark matter ($\eta$ADM) 
or simply as $\eta-$asymmetric dark matter. 

Several experiments committed to  direct dark matter searches 
show evidence of positive particle detection,  although  these results 
are still very controversial and not universally accepted:
the DAMA/LIBRA \citep{2012PhRvL.108e1302B,Bernabei:2008ik} and CoGeNT \citep{2011PhRvL.106m1301A} 
experiments find evidence of a particle candidate  with a 
mass of the order of a few GeV (likely between  $5$ and $12$ GeV),  and a
spin-independent scattering cross section of baryons of the order of $10^{-40}\;{\rm  cm^2}$.
The CRESS \citep{2012PhRvD..85b1301B} experiment also points to  unexplained events consistent with  direct detection of a light mass particle. 
Unfortunately,  other direct detection experiments, such as  CDMS \citep{2011PhRvL.106m1302A} and 
 XENON10/100 \citep{2011PhRvL.107m1302A,2011PhRvL.107e1301A}  find no indication of
 the existence of such particles. Nevertheless, several theoretical explanations based on the existence of asymmetric dark matter 
 have been suggested to explain and accommodate all these positive and negative detections 
\citep[e.g.][]{2010JCAP...08..018C,2011JCAP...11..010F,2011PhRvD..84h3001H,2011PhRvD..84b7301D}.

In this paper, we investigate the origin of such light $\eta-$asymmetric dark matter particles, 
and discuss how this type of matter influences the evolution of stars.
We  look for their  impact on the present structure of the Sun. 
The Sun, by means of two groups of observables, helioseismology 
and solar neutrino fluxes, provides one of the most powerful tests of  stellar evolution  \citep{2011RPPh...74h6901T} and
of alternative theories of modern physics and cosmology \citep{2012ApJ...745...15C,2012ApJ...746L..12T,2002MNRAS.331..361L}.
Although both groups of observables are equally relevant for such types of studies, 
we have chosen  preferentially to study the impact on solar neutrino fluxes, 
as neutrinos are the most sensitive probe of changes in the structure of the solar core.
In particular, we will compare the neutrino fluxes of a "Sun" evolving in a halo of asymmetric dark matter
with  the current measurements  of solar neutrino fluxes, namely, the $^8B$ and $^7Be$ neutrino fluxes.
Moreover, we will complete the study with a succinct helioseismology diagnostic.

In the reminder of this paper,  we  consider that the dark matter asymmetry 
is identical to the baryon asymmetry, and 
likewise leads to an unbalanced amount of particles and antiparticles.
This $\eta-$asymmetry occurs well before the epoch of thermal decoupling of the  dark matter. 
We do not discuss here any of the possible mechanisms for the generation 
of such asymmetry, but rather treat this quantity as a free parameter. 
This asymmetric dark matter framework is used to study the impact of such matter 
in the evolution of the Sun for a  wide range of particle masses, 
annihilation cross sections and dark matter asymmetries.

In Sec. 2 we present the basic  properties  of $\eta-$parametrised asymmetric dark matter
and explain how the relic dark matter density of the present day Universe is computed.
In Sec.(s) 3, 4 and 5  we discuss the changes caused in the Sun  by the presence of $\eta-$parametrised 
asymmetric dark matter in its core, as well as the impact on  the flux of solar neutrinos and helioseismology. 
In the last section, we summarize our results and draw some conclusions.

 \begin{figure*}[ht]
 \centering
 \plottwo{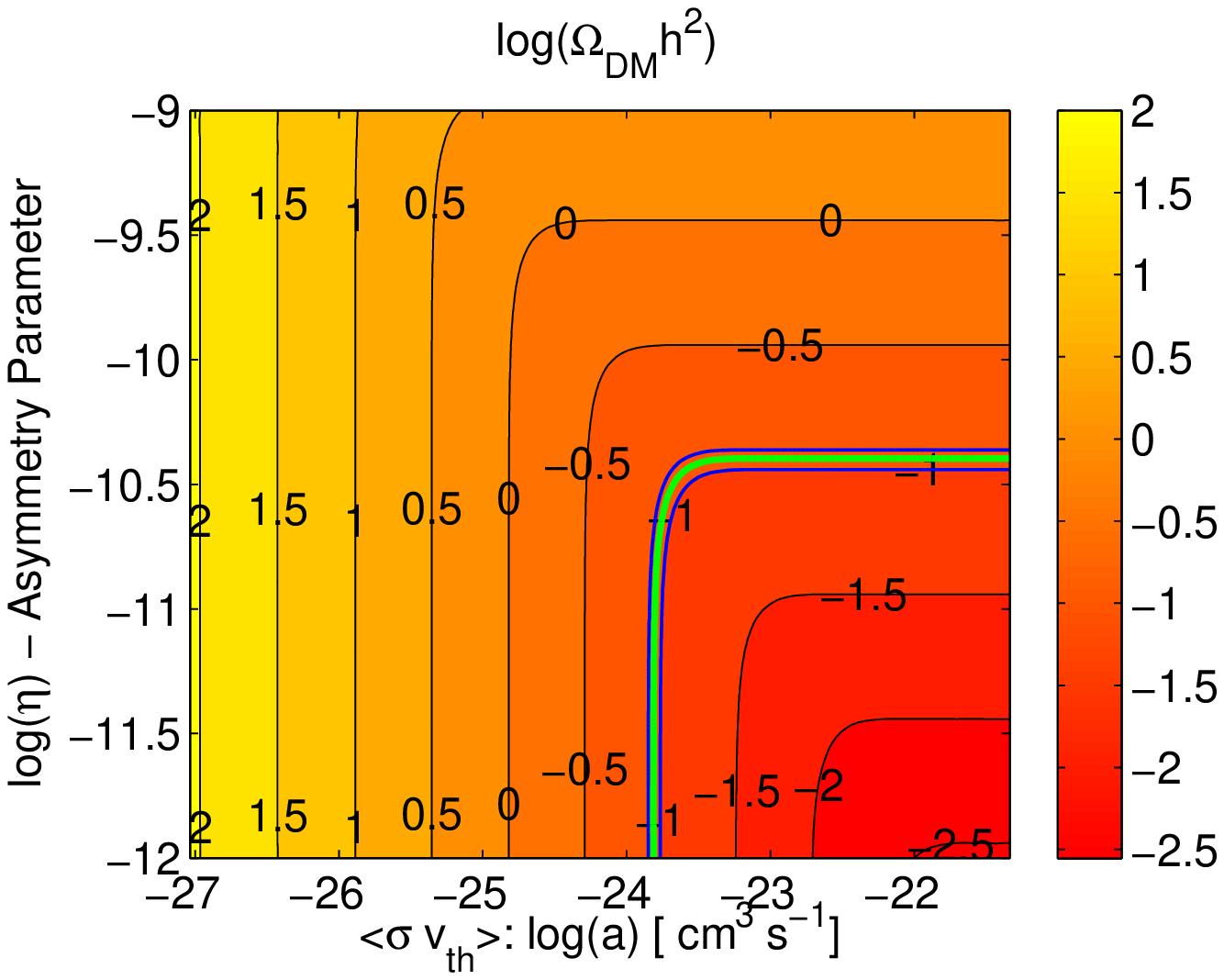}{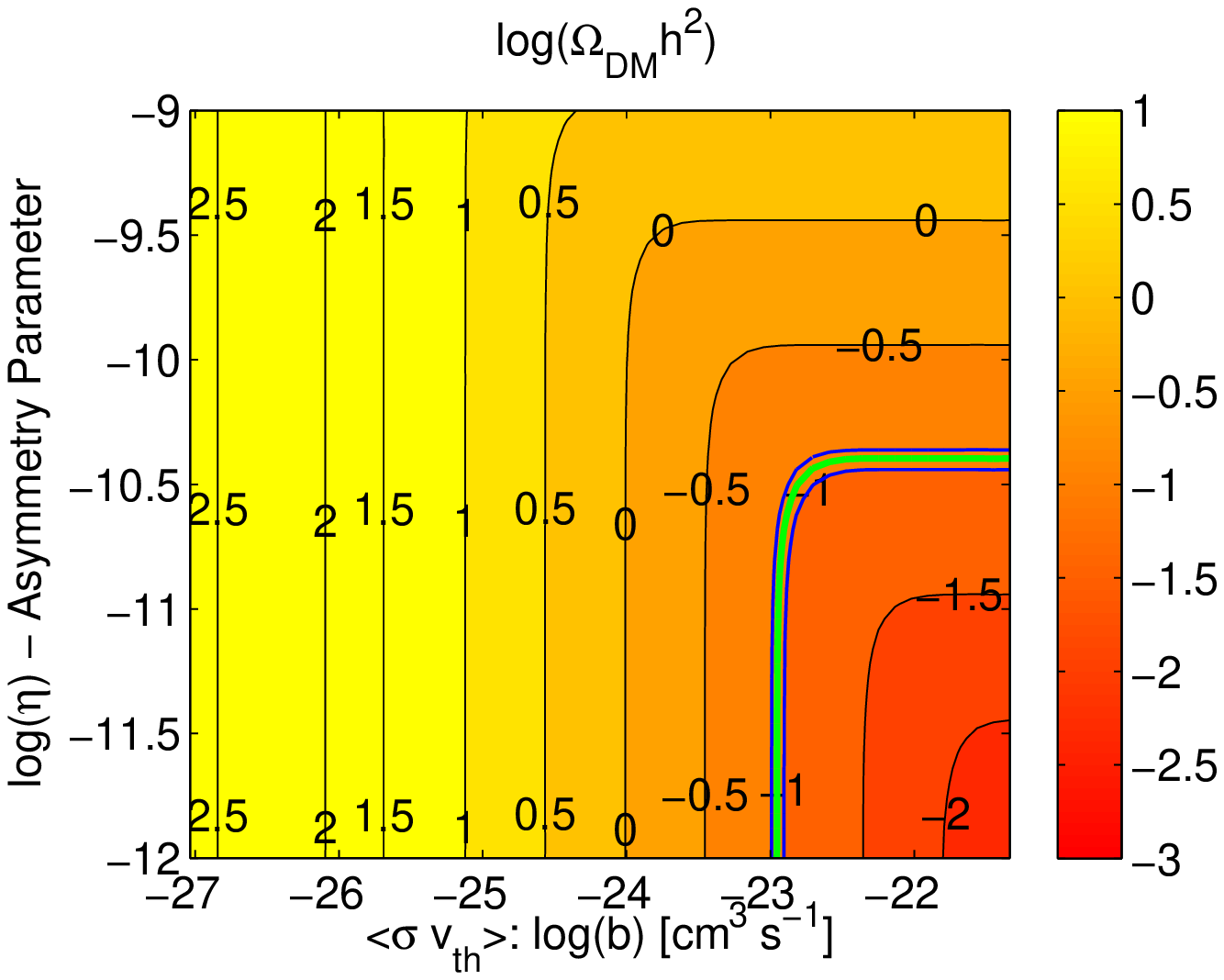}
 \label{fig:1}
 \caption{Asymmetric dark matter particle models: The figure shows iso-curves of the relic dark matter density  $\Omega_{DM}h^2$ 
 as a function of the asymmetry parameter $\eta$ and the annihilation cross section $\langle \sigma v\rangle $:
 (a)  s-wave annihilation channel ($a\ne 0$,$b=0$)  and (b) p-wave annihilation channel ($a= 0$,$b\ne 0$).
 The results are for dark matter particles with $m_\chi = 10 \;{\rm GeV}$, $g_\chi = 2$ and $g_\star = 90$ (see text).
 In the figure, indicated with blue lines are the $\eta-$asymmetric dark matter particle cosmological models compatible with the present day dark matter density. 
 In the figure, indicated with blue lines are the asymmetric dark matter particle cosmological models 
 compatible with the present day dark matter density. This cosmological model is computed for  two types of annihilation channels, $(a,\eta)$ or $(b,\eta)$
 (see text). We consider the observational  window 
 to be such that $0.10 \le\Omega_{DM}h^2 \le 0.12 $ (or $ -1.0 \le \log{(\Omega_{DM}h^2)} \le  -0.92$).
 This corresponds to the region between the blue lines. 
The observed  value of $\Omega_{DM}h^2$is shown as a green line.
The current measurements of the cosmological density and baryonic asymmetry are $\Omega_{DM}h^2 =0.1109\pm 0.0056$
and $\eta_{B}=(0.88\pm 0.021)\times 10^{-10}$,  or $\log{(\Omega_{DM}h^2)}\sim -0.9551$ 
and $\log{(\eta_B)}\sim -10.06$ \citep{Larson:2011ji,2011A&A...535A..26H}. }
 \end{figure*}

\section{Asymmetric dark matter in the early Universe}  

In the early stages of the Universe, the cosmic fluid rapidly reached thermal equilibrium,
as a consequence of  standard and dark matter particles having a very high rate of interactions. 
In this work, we allow dark matter to be constituted by a 
mixture of particles and antiparticles, $\chi$ and $\bar{\chi}$ with 
a mass  $m_\chi$,  and $g_\chi$ being the number of internal degrees of freedom \citep{2006PhRvD..73l3502D}.
The Universe is considered to have a temperature $T$ and an  
effective number of relativistic degrees of freedom $g_\star$ \citep{2011JCAP...07..003I}.  
Unlike symmetric dark matter,  $\chi$ and $\bar{\chi}$ particles  are not necessarily identical.
Following the usual convention, the asymmetric nature of the dark matter is defined by the
parameter $\eta$ which is equal to the difference of particle and antiparticle populations.
Without loss of generality, we conventionally define  $\eta \ge 0$, i.e., particles are more abundant than anti-particles. Here
$\eta $ stays constant  throughout the evolution of the Universe. 
The asymmetry $\eta$  is similar to  the asymmetry for baryons $\eta_{B}$,
both of which  originate in the early phases of the Universe. 
If not stated otherwise, in the numerical examples, the population of  
$\chi$ and $\bar{\chi}$ particles is constituted by elementary particles with mass $m_\chi \sim {\rm 10\; GeV}$, 
$g_\star = 90$ and $g_\chi=2$. This is equivalent to a particle population formed by Dirac fermions   
\citep{2010PhLB..687..275D,2006PhRvD..73l3502D}. 

As the Universe expands, the interactions between all particles become sparse,
the temperature of the plasma drops, the Universe gets cooler, 
up to the moment that  the $\chi\bar{\chi}$ annihilation rate drops below the Hubble expansion rate, 
leading to dark matter becoming decoupled from the rest of the cosmic fluid.
The dark matter density, $\Omega_{DM}$ , has frozen out and has been constant ever since.
The relic densities of $\chi$ and $\bar{\chi}$ particles 
are determined by  solving the coupled system of Boltzmann equations that follows
the time evolution of the number density of particles and antiparticles in the expanding universe
 \citep{2011JCAP...07..003I}.
 
The present-day dark matter density, $\Omega_{DM}$, depends
on  the relic  densities of particles  and antiparticles, $\Omega_{\chi}$ and $\Omega_{\bar{\chi}}$, 
which in turn depend on the mass of the particle, $m_\chi$,  the
 $\chi\bar{\chi}$ annihilation cross section,  $\sigma$
and the dark matter asymmetry $\eta$.

Primarily,  the value of $\Omega_{DM}$  relies on  the properties of the $\chi\bar{\chi}$ annihilation cross section.  
In most cases, the thermal averaging of the $\chi\bar{\chi}$ annihilation cross section times  the relative velocity of colliding particles $v$ can be expanded as
$\langle\sigma v\rangle = a +b v^2 + {\cal O} (v^4)$.  If the annihilation initial state is {\it unsuppressed}, 
 the first term of the previous equation dominates ($a\ne 0$). This process is known as the s-wave annihilation channel. 
Alternatively, if the initial state is {\it suppressed}, $ a = 0$ and $b \ne  0$, this is the  p-wave annihilation channel.
This expansion is valid for all known examples of  s-wave channel, p-wave  channel or both channels,
 up to an accuracy of a few percent.  
The expression $\langle\sigma v\rangle$ can be simplified further, if  the reacting particles are non-relativistic,
$\langle\sigma v\rangle = a +6 b x^{-1} + {\cal O} (x^{-2})$ where $x$ 
defines the ratio of  $m_\chi$ over the temperature of the Universe $T$ \citep{2011JCAP...07..003I}.

In the case of symmetric dark matter particles ($\bar{\chi}=\chi$),  the  relic abundance 
$\Omega_{\chi}$ is mainly determined by the annihilation rate $\langle\sigma v\rangle$. Furthermore,  $\Omega_{\chi}$
is equal to the dark matter value $\Omega_{DM}$.  Alternatively, in the case of asymmetric dark matter ($\bar{\chi}\ne\chi$), 
$\chi$ and $\bar{\chi}$  particle contributions have to be added such  that $\Omega_{DM}= \Omega_{\chi}+\Omega_{\bar{\chi}}$, 
as there are more particles than antiparticles  (or the reverse). The antiparticles are annihilated away more efficiently, 
with large numbers of particles left behind without a partner to annihilate them. Consequently, the relic dark matter abundance is determined not only by the $\langle \sigma v \rangle$  as in the symmetric dark matter case,  but also by the asymmetry parameter  $\eta$.
 
 .
 
With the objective of computing the evolution of the Sun in different cosmological scenarios, we start by determining the relic densities of particles and antiparticles, $\Omega_{\chi}$ and $\Omega_{\bar{\chi}}$ for specific $\eta-$asymmetric dark matter models. The  computation is done by following closely the numerical procedure of  \citet{2011JCAP...07..003I}. 
Figure 1 shows the dark matter density $\Omega_{DM}$ for 
light asymmetric particles with $m_\chi \sim 10\; GeV$  computed for
several values of $\eta$ and $\langle \sigma v \rangle$.  
Two  types of $\chi\bar{\chi}$ annihilation rates  are considered,  pure s-wave and a pure p-wave annihilation channels.
Annihilation channels with non-vanishing $a$ and $b$ terms are qualitatively similar to the previous ones.  

In general, such results show that there are two limiting asymmetric dark matter scenarios, 
one for very high  and other for  very low values of $\eta$:
(i) for the high values of $\eta$, the dark matter is "strongly" asymmetric as the relic density  
is dominated  by particles over antiparticles, $\Omega_{\bar{\chi}}\ll \Omega_{\chi} \approx \Omega_{DM} $;
this scenario is identical to the baryogenesis process related to ordinary baryons.
(ii) for the low values of $\eta$, the dark matter is "weakly" asymmetric or symmetric ($\eta\approx 0$),  
$\Omega_{\bar{\chi}}\approx \Omega_{\chi} \approx \Omega_{DM} $; the final relic abundances 
of particles  and antiparticles are comparable. 
This generally corresponds to the standard thermal WIMP scenario.

If we consider the present measurements of $\Omega_{DM}$ \citep{2011A&A...535A..26H}, we find that  
the number of asymmetric dark matter scenarios that are compatible with 
observations  is relatively reduced, i.e.,   
only $\eta-\langle \sigma v \rangle$  dark matter models with a relic density 
$0.1053 \le \Omega_{DM}h^2\le 0.1165$. 
Nevertheless, for reasons of convenience\footnote{The recent measurements of  $\Omega_{DM} h^2$ are such that 
$0.1053 \le \Omega_{DM}h^2\le 0.1165$ or $-0.9776 \le \log{(\Omega_{DM}h^2)} \le -0.9337$, nevertheless for clarity of argument
we choose a slightly larger interval of $\Omega_{DM} h^2$ which includes the observational window.}, 
in this work we choose the observational interval to be such that  $0.10 \le \Omega_{DM}h^2\le 0.12$
(region between blue lines in Figure 1).

If we restrict our analysis to the  asymmetric models of particles with a mass of 10 GeV  
compatible with  dark matter density observations, it is possible to determine a critical value of 
$\langle \sigma v \rangle_c$ for which there is a change of dark matter regime (see figure 1). The critical value $\langle \sigma v \rangle_c$ 
is approximately $1.6\times 10^{-24}\;{ \rm cm^3 s^{-1}}$ and $1.2 \times 10^{-23}\;{cm^3 s^{-1}}$ for s-wave and p-wave annihilation channels.
There is also a maximum value of asymmetry  $\eta_c$, above which the asymmetric dark matter models 
have a  $\Omega_{DM}$ larger than the current observational value.  $\eta_c$  is equal to $3.6\times 10^{-11}$ for the s-wave annihilation channel and 
$2\times 10^{-11}$  for the p-wave annihilation channel.
If a model has $\langle \sigma v \rangle < \langle \sigma v \rangle_c$, $\Omega_{DM}$ 
becomes uniquely dependent on $\eta$ (independent of $\langle \sigma v \rangle$ ). 
This corresponds to cosmological models for which 
the $\chi\bar{\chi}$ annihilation rate is so efficient that the relic abundance depends
only on the  initial $\eta$ asymmetry. 
Conversely, if a model has $\langle \sigma v \rangle > \langle \sigma v \rangle_c$,
the dark matter density becomes purely determined by the value of $\langle \sigma v \rangle$ 
(independent of $\eta$). It follows that the dark matter density content is determined 
by the low value of the annihilation rate.

It is worth noticing that the dependence of $\Omega_{DM}$  with the 
particle mass is quite small for asymmetric dark matter models of light particles ($m_{\chi} \le 20\; {\rm GeV}$). 
In cosmological dark matter models for particles with a mass between 5 GeV  and 20 GeV,  
$\langle \sigma v \rangle_c $  for both channels is identical to the case of a particle with a mass of 10 GeV.  
For the same range of masses, $\eta_c$  varies between  $8-2\times 10^{-11} $ for a
s-wave ($a\ne 0$) channel and  $7-2 \times 10^{-11}$ for p-wave channel ($b\ne 0$). 

\begin{figure*}[ht]
\centering
\plottwo{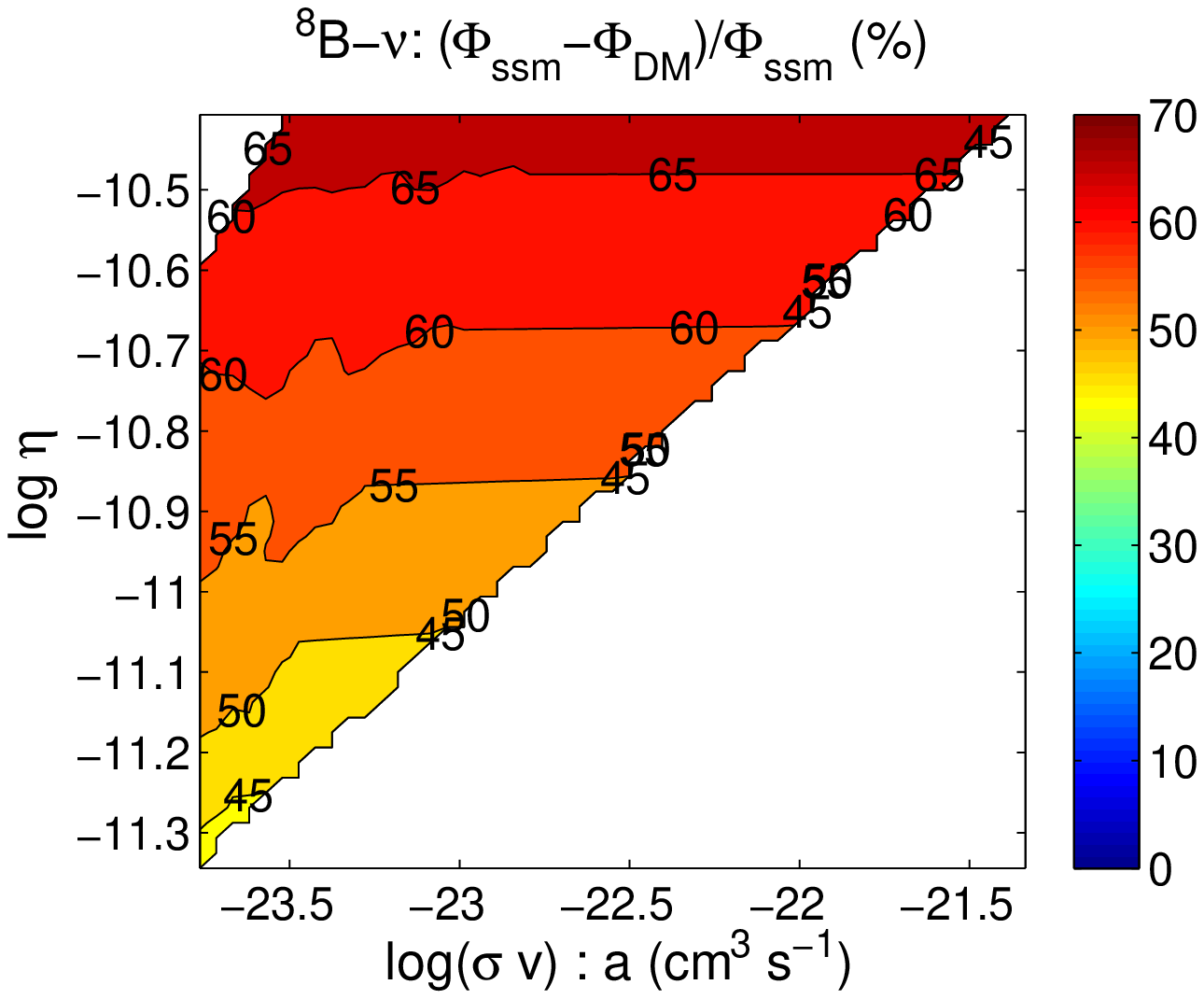}{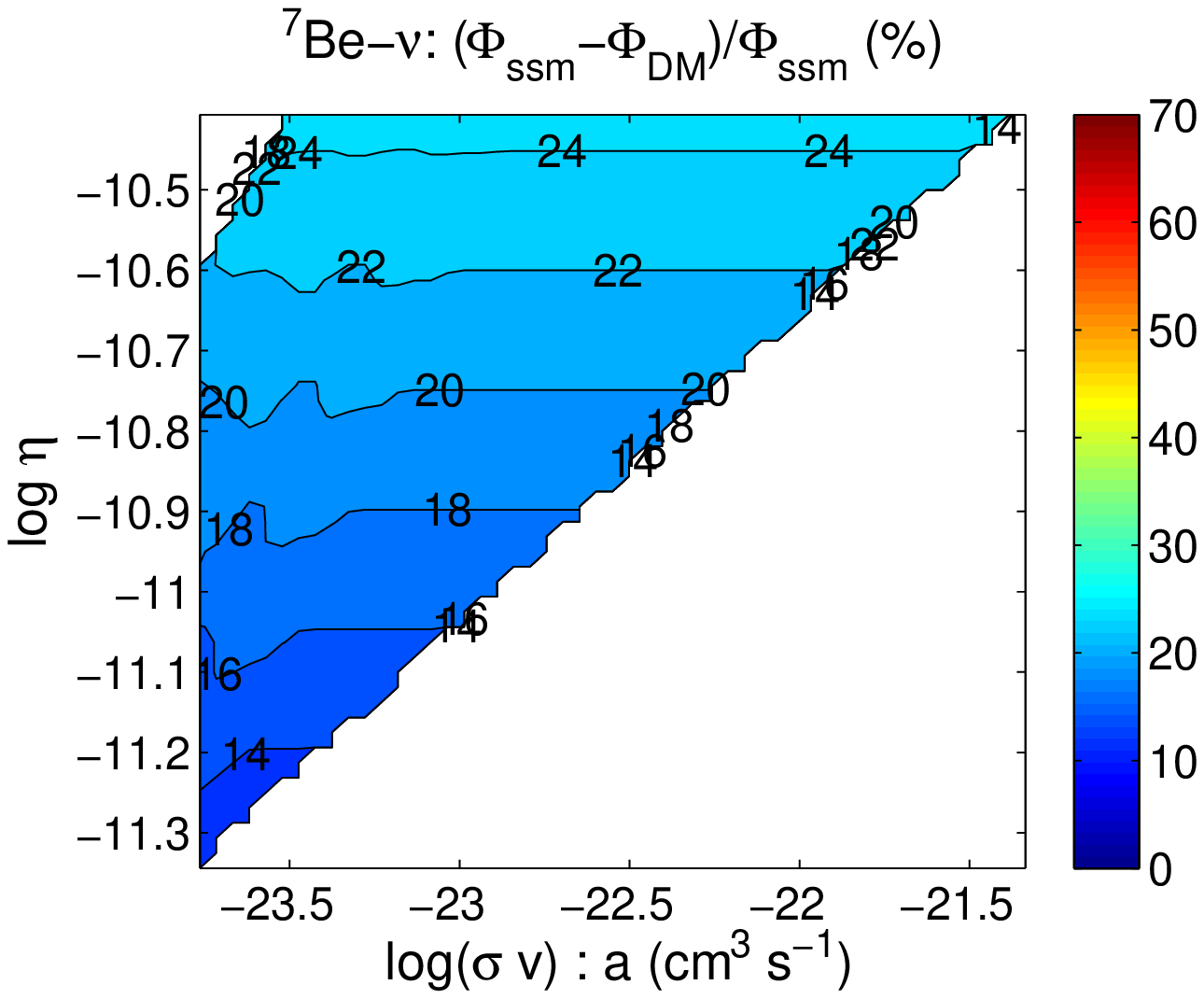}\\
\plottwo{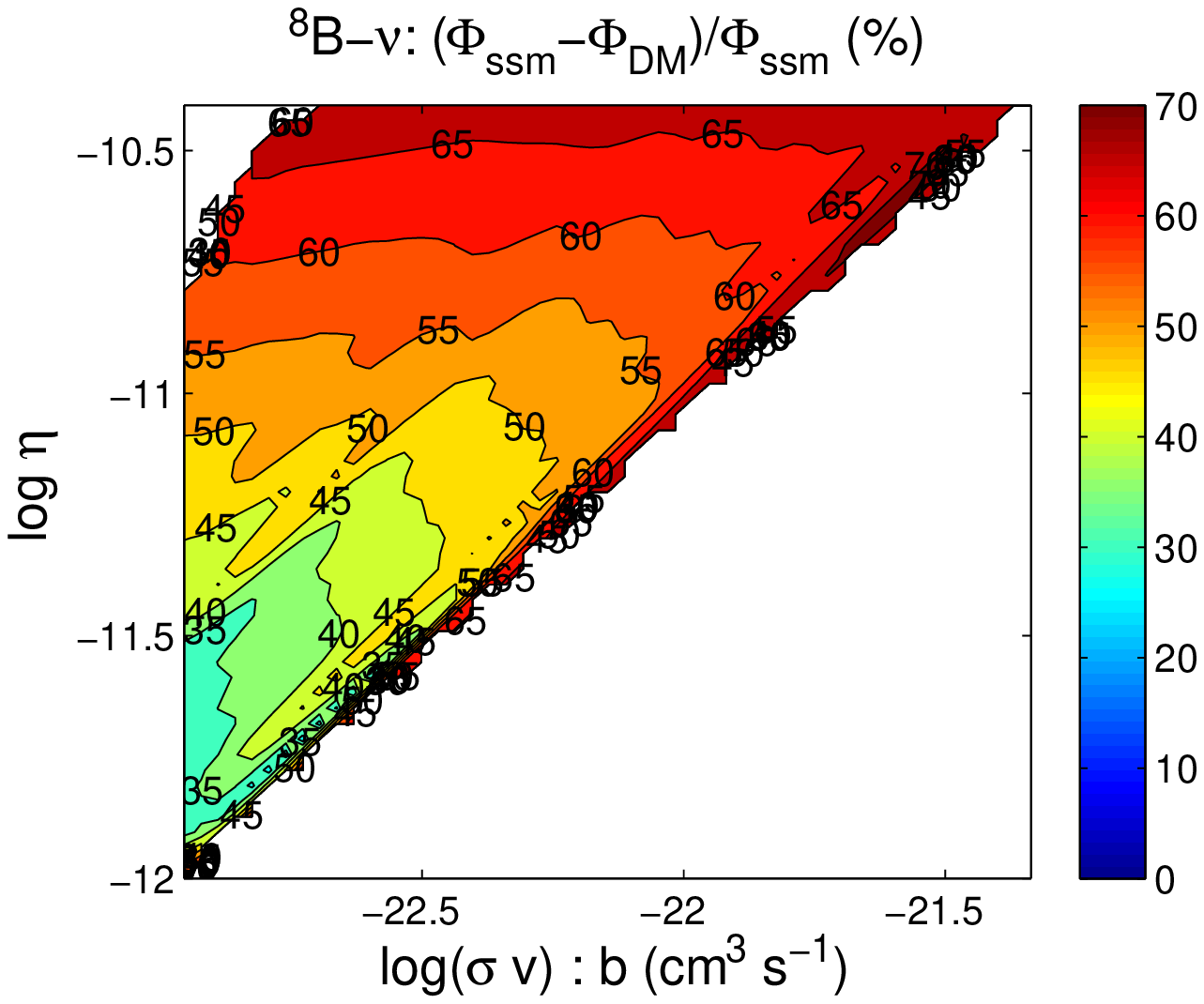}{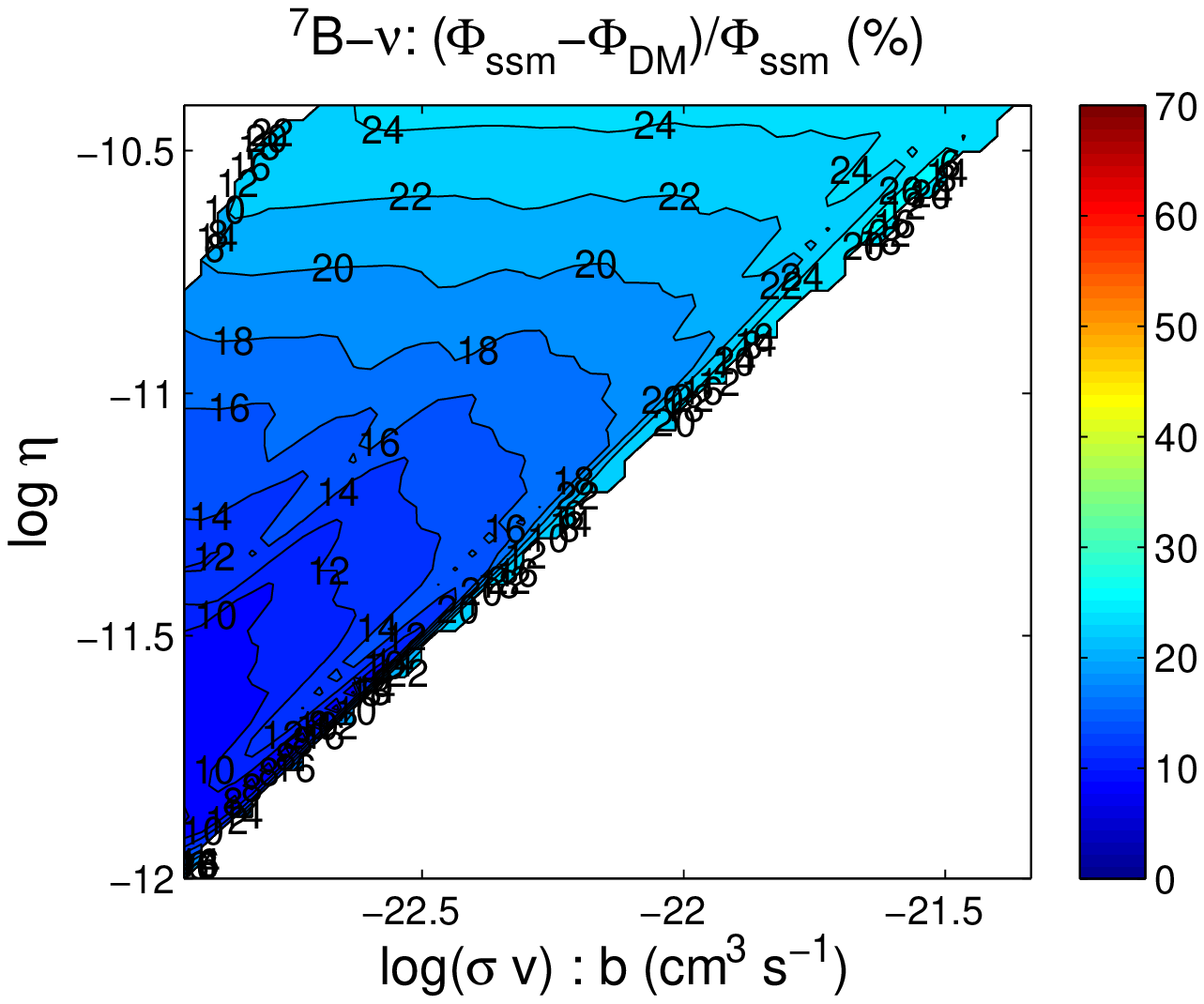}
\caption{ Asymmetric dark matter model impact on the solar neutrino fluxes:
The asymmetric dark matter particles have a mass of $10\;{\rm GeV}$ and 
a scattering cross section of baryons such that $\sigma_{SD}=10^{-40} \; {\rm cm^2}$ and $\sigma_{SI}=10^{-36} \; {\rm cm^2}$.
The asymmetric dark matter cosmological models have a value of $\langle \sigma v \rangle$
and $\eta$  for which the cosmological models have  $\Omega_{DM} h^2$ value consistent with the observational window, i.e., 
$0.10 \le \Omega_{DM}h^2\le 0.12$ or $ -1.0 \le \log{(\Omega_{DM}h^2)} \le  -0.92(08) $. This corresponds to asymmetric dark
matter models for which $\Omega_{DM}h^2$ is in between the blue lines Figure 1 (see text).}
\end{figure*}

\section{Capture of $\eta-$parametrised asymmetric dark matter by the Sun}  
 
The accumulation of particles ($\chi$) and antiparticles ($\bar{\chi}$) 
inside the star depends on the star's gravitational field 
and the interaction of the dark matter particles with baryons.
In our description of the impact of the $\eta-$asymmetric dark matter
on  stellar evolution, we follow closely earlier discussions of
the impact of "classical" asymmetric dark matter\footnote{In the literature, we find another description of asymmetric dark matter,that is distinct of the type of dark matter discussed in this work. Although both types of asymmetric dark matter accumulate by identical processes inside the star, their effects in the evolution of stars is rather different. We decided to call $\eta-$asymmetric dark matter to highlight the difference} with the evolution of the Sun and other stars 
\citep{2009MNRAS.394...82S,2010PhRvD..82j3503C,2010PhRvD..82h3509T,2009ApJ...705..135C,2011PhRvD..83f3521L}.

In the following we discuss the capture and accumulation of $\eta-$parametrised asymmetric dark matter by the Sun. Although, such particles interact with baryons in a similar way as WIMPs, the fact that we have two distinct populations of particles,  leads to a quite different impact on the evolution of the star. Like WIMPs, the amount of  $\eta-$asymmetric dark matter captured by the star depends explicitly on the mass of the dark matter particle $m_\chi$, the cross-section for scattering with baryons, namely, the spin-dependent scattering cross-section $\sigma_{SD}$, and the spin-independent scattering cross-section $\sigma_{SI}$  \citep{2011PhRvD..83f3521L}. However, unlike WIMPs, $\eta-$asymmetric dark matter, depends also on the two channels (p-wave and s-wave) for annihilation   $\langle\sigma v\rangle$ and the $\eta-$asymmetry parameter. The total number of particles $N_\chi$ and antiparticles $N_{\bar{\chi}}$ that accumulates inside the Sun at a certain epoch is computed by solving the system of coupled equations:
 
\begin{eqnarray}
\frac{dN_{i}}{dt}=C_{i}-C_{a} N_{\chi}N_{\bar{\chi}},
\label{eq-Ni}
\end{eqnarray}
with $i$ being $\chi$ or  $\bar{\chi}$. The constant $C_{i}$ gives the rate of capture of 
particles (antiparticles) from the dark matter halo, 
and  $C_{a}$ gives the annihilation rate of particles and antiparticles in the Sun. 
This approach is entirely different from other studies of accumulation of dark matter inside stars,
in part because previous work only studied the capture of the population $N_\chi$ of dark matter particles.
In such cases, the system of coupled equations (\ref{eq-Ni}) is substituted by one single equation. 
In the following, we will consider dark matter particles with a mass larger than $5\; GeV$,
for which  the evaporation of particles is negligible \citep{1990ApJ...356..302G}.
Furthermore, we neglect the rate of capture of dark matter particles by scattering off 
other dark matter particles that have already been captured within the Sun  
\citep{2009PhRvD..80f3501Z}.

The capture rate for particles and antiparticles by the Sun's gravitational field and their scatterings off  baryons
is computed numerically from the integral  
expression of \citet{1987ApJ...321..571G},  implemented as indicated in \citet{2004JCAP...07..008G}.
The description of how this capture process is implemented in our code is discussed  
in  \citet{2011PhRvD..83f3521L}. The scattering of particles and antiparticles 
on  baryon nuclei are identical.  It follows that, similarly to the symmetric dark matter case, $\sigma_{SD}$ 
is only relevant for hydrogen nuclei; and $\sigma_{SI}$
is important for the scattering of dark matter particles on heavy nuclei.  
If the value of  $\sigma_{SI}$ is larger or equal to  $\sigma_{SD}$,
the capture of dark matter particles is dominated by the collisions 
with heavy nuclei, rather than by  collisions with hydrogen \citep{2011PhRvD..83f3521L}.

If not stated otherwise,  we assume that the density of dark matter in the 
halo is $0.38 \; {\rm GeV cm^{-3}}$\citep{2010JCAP...08..004C}, 
the amount of particles and antiparticles in the halo is in the same proportion as in the local Universe,
 the stellar velocity of the Sun is $220 \; {\rm km s^{-1}}$ 
and the Maxwellian velocity dispersion  of dark matter particles is $270 \; {\rm km s^{-1}}$ 
\citep[e.g.,][]{2005PhR...405..279B}. 
Recent measurements of local dark matter density yet to be published
propose a value of $0.3\; {\rm GeV cm^{-3}}$ \citep{2012arXiv1205.4033B} and 
$0.85\; {\rm GeV cm^{-3}}$ \citep{2012MNRAS.tmp.3493G}. 
The annihilation rate $C_{a}$ is computed from
\begin{eqnarray}
 C_{a}=N_{\chi}^{-1}N_{\bar{\chi}}^{-1}\;\int \langle\sigma v\rangle (r)\;   n_{\chi}(r) n_{\bar{\chi}} (r)\;4\pi^2 r^2 \;dr, 
\end{eqnarray} 
where $ \langle\sigma v\rangle (r) $ is the $\chi\bar{\chi}$ annihilation rate as defined in the previous section,
and  $n_{\chi}(r) $ and  $n_{\bar{\chi}}(r)$ are the number density of particles and antiparticles.

In this paper, in contrast to previous work, we assume that the dark matter particles and antiparticles present inside the Sun annihilate by the same physical process as the one that occurs in the early Universe, therefore, i.e.,
$\langle\sigma v\rangle = a +6 b x^{-1} + {\cal O} (x^{-2})$  where $x$ is now the ratio of mass of the dark matter particle over the local temperature of the Sun's plasma.
It follows that inside the Sun
the  s-wave ($a\ne 0$) annihilation channel  is identical to the one present in the early Universe,
but the p-wave ($b\ne 0$)  annihilation channel is quite different, as the temperature inside the Sun
is larger than the temperature in the  early Universe. 

As the star evolves, after some time,  $t$ is larger than $\tau_{A}=\sqrt{C_{a}C_{\chi}}$,  
the rate of accumulation of particles inside the Sun $\dot{N}_{\chi}$ is proportional to the 
difference between the capture rate of the particles and antiparticles, i.e., $\dot{N}_{\chi} = D_{\chi}$
with  $D_{\chi}=C_{\chi}- C_{\bar{\chi}} $. It follows that the total number of particles and antiparticles
is given by $N_{\chi}\approx N_{\bar{\chi}}+ D_{\chi}\; t$ and
$N_{\bar{\chi}}\approx C_{\bar{\chi}}/(C_{a}  D_{\chi} \; t ) $ \citep{1987NuPhB.283..681G}.
The value of $\tau_{A}$ depends of the values of dark matter parameters.  For a typical dark matter 
halo of density  $0.38 \; {\rm GeV cm^{-3}}$ constituted by particles with a mass of $10 {\rm GeV }$ 
and annihilation rate  $10^{-24}{\rm cm^{3} s^{-1} } $,   $\tau_{A}$ of the order of 1 million year.
Particles and antiparticles, once captured by the Sun during its evolution, end up in thermal equilibrium with baryons.
The particle (and antiparticle) population  follows a Maxwellian velocity distribution. The number density
of particles (antiparticles), $n_{i}(r)\approx N_{i}e^{-m_\chi \phi(r) /T_{c} }$  where $T_c$ in the central temperature of the star and
$\phi(r)$ is the gravitational potential \citep{2009ApJ...705..135C,GiraudHeraud:1990uu}.

\begin{figure*}[ht]
\centering
\plottwo{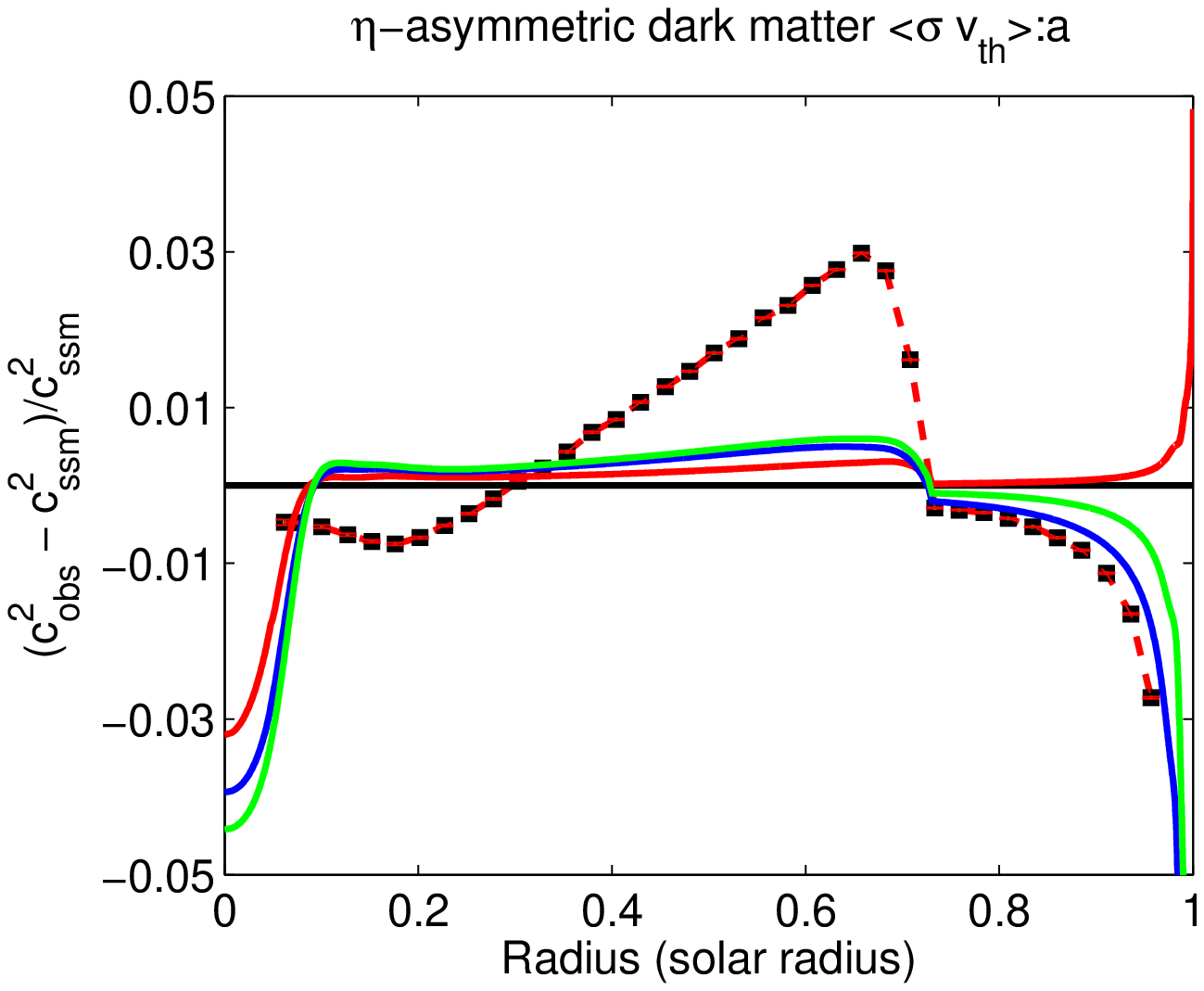}{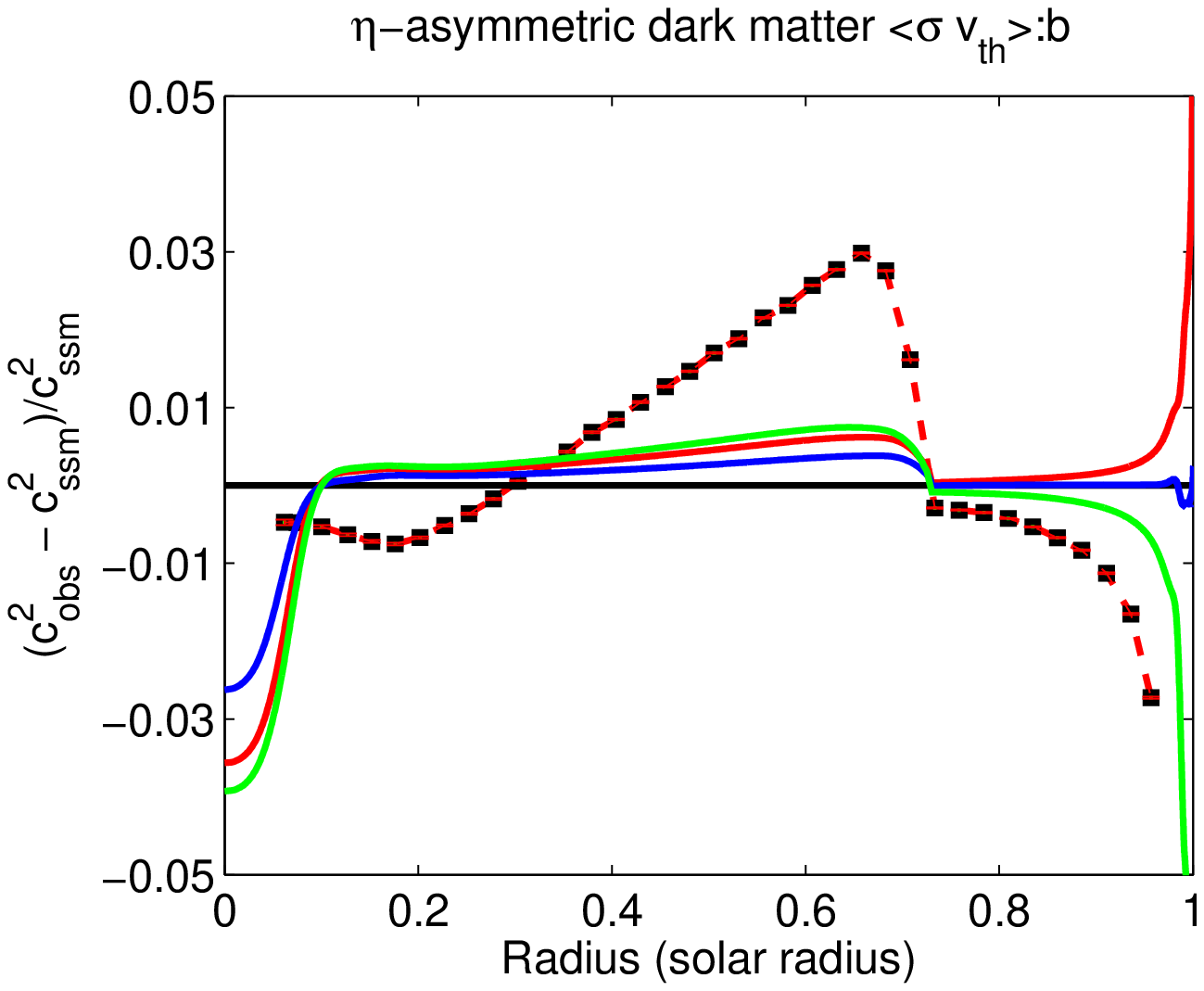}
\caption{
Comparison of the sound speed radial profile between the solar standard model and different solar models evolved within $\eta-$asymmetric dark matter halos: The red dotted curve corresponds to the difference between inverted profiles and our solar standard model \citep{2009ApJ...699.1403B,1997SoPh..175..247T}. The dark matter particles have a mass of 10 GeV and a scattering cross section oﬀ  baryons such that $\sigma_{SD}= 10^{-40} \; {\rm cm}^2 $  and  $\sigma_{SI}=10^{-36} \; {\rm cm}^2 $. The annihilation dark matter particles in the early Universe and inside the Sun have an annihilation $\langle\sigma v\rangle$  such 
$\langle\sigma v\rangle=a+b v^2 $ (see main text for details). The $\eta-$asymmetric dark matter particles have the following properties: 
(i) {\bf Left figure}, s-wave annihilation channel $a \ne 0, b = 0$, 
(a) $\rho_\chi =0.21\; {\rm GeV cm^{-3}}$ and $a =1.72\; 10^{-24} \;{\rm cm^3 s^{-1}}$ (red curve);
(b) $\rho_\chi =0.34\; {\rm GeV cm^{-3}}$ and $a =2.25\; 10^{-24} \;{\rm cm^3 s^{-1}}$ (blue curve)
(c) $\rho_\chi =0.38\; {\rm GeV cm^{-3}}$ and $a =4.61\; 10^{-24} \; {\rm cm^3 s^{-1}}$ (green curve)
(ii) {\bf Right figure}, p-wave annihilation channel $a = 0, b \ne 0$, 
(a) $\rho_\chi =0.19 \; {\rm GeV cm^{-3}}$ and $b =1.11\; 10^{-23} \; {\rm cm^3 s^{-1}}$ (red curve);
(b) $\rho_\chi =0.22 \; {\rm GeV cm^{-3}}$ and $b =1.11\; 10^{-23} \;{\rm  cm^3 s^{-1}}$ (blue curve)
(c) $\rho_\chi =0.38\; {\rm GeV cm^{-3}} $ and $b =4.61\; 10^{-22} \; {\rm cm^3 s^{-1}}$ (green curve)
The dark matter halo has a total dark matter density, $\rho_{DM} =0.38 \; {\rm GeV cm^{-3}} $, such that $\rho_{DM} = \rho_{\chi}  + \rho_{\bar{\chi}}$ The density of particles and antiparticles, $\rho_{\chi}$ and $\rho_{\bar{\chi}}$  are proportional to $\Omega_{\chi}$  and $\Omega_{\bar{\chi}}$.}
\end{figure*} 

\section{Impact of $\eta-$parametrised asymmetric dark matter on the Sun's evolution}  
 
The code used to compute the evolution of asymmetric dark matter is 
a modified version of the stellar evolution code CESAM 
\citep{1997A&AS..124..597M}. The basic reference model without dark matter
is calibrated to produce a solar standard model \citep{1993ApJ...408..347T}  identical to others in the
literature \citep{2010ApJ...713.1108G,2009ApJ...705L.123S,2005ApJ...621L..85B,2010ApJ...715.1539T}.
The microscopic physics
(updated equation of state, opacities, nuclear reactions rates, 
and an accurate treatment of microscopic diffusion of heavy elements)
are in full agreement with the standard picture. 
The solar mixture of \citet{Asplund:2005uv} is used in the computation of models.
The modified evolution models are calibrated to the present solar radius, 
luminosity, mass, and age  \citep[e.g.][]{2011RPPh...74h6901T}. 
The models are required to have a fixed value of the photospheric ratio 
$(Z/X)_\odot$, where  X and Z are the mass fraction of hydrogen and the mass fraction of elements is heavier than helium.
The value of $(Z/X)_\odot$ is determined according to the solar mixture \citet{Asplund:2005uv}.

The  star evolves from the beginning of the pre-main sequence
until its present age. Each solar model has more than 2000 layers,
and it takes more than 80 time steps to arrive to the present age.
For each set of dark matter parameters, a solar-like calibrated model is obtained by
automatically adjusting the helium abundance and the convection mixing length parameter until the total luminosity
and the solar radius are within $10^{-5}$ of the present solar values. Typically a calibrated solar model
is obtained after a sequence of  10 intermediate solar models, although the models with a large concentration of dark matter 
need more than 20 intermediate models.
 The increase of the number of iterations is related  with the rapid variation of the  structure
in the center of the Sun. The values of the calibrating parameters,  mixing-length parameter and
 the initial content of Helium changes slightly relatively to the standard solar model, 
depending upon the accumulation of dark matter in the core of the star.  The maximum variation 
obtained for these parameters due to the accumulation of dark matter in the centre of the Sun 
is an increase of the mixing-length parameter of 5\% and the initial content of Helium is decreased by 0.5\%.

The  $\eta-$asymmetric dark matter particles impact the evolution of the Sun,
by a physical process  identical to  WIMPs. Likewise they provide an effective mechanism for the transport of energy,
for which the efficiency  depends locally on the average mean free path of 
the dark matter particles between successive collisions with baryons.
Two distinct regimes of the transport of energy are usually considered:
local  and non-local transport corresponding to small and large average mean free paths. 
Both energy transport mechanisms are implemented in our code \citep{2011PhRvD..83f3521L}.
This physical process leads to the 
reduction of the temperature gradient \citep{2002MNRAS.331..361L,2010ApJ...722L..95L}.
In the more extreme case, 
the high frequency of collisions between baryons and dark matter particles
forms an isothermal core \citep{2002PhRvL..88o1303L}.

This new type of asymmetric dark matter can have a much larger effect on the evolution of the
Sun than the usual symmetric dark matter \citep{2010ApJ...722L..95L}, 
because the star can  accumulate a much larger amount of
particles (and antiparticles) in its core than in the latter case. This is due to the fact that asymmetric dark matter 
depends, among other parameters, on  two major parameters that determine the concentration 
of dark matter inside  the star: dark matter asymmetry, which determines 
the unbalanced amount of particles and antiparticles;  and the annihilation cross-section, which
establishes the annihilation efficiency of particles and antiparticles. 
This is consistent with the results shown in Figure 1.
 
Figure 2 shows the variation for the $^8B$ and $^7Be$ neutrino fluxes 
relative to the solar standard models.  The large reduction in  $^8B$ and $^7Be$ neutrino fluxes 
is a direct consequence of the reduction of the temperature inside the solar core. 

This diminution of temperature is also visible in the sound speed profile near the centre of the Sun. Figure 3 shows the profile of sound speed computed for a few $\eta-$asymmetric dark matter scenarios. It also shows the observed sound speed profile obtain from the helioseismology data of the BISON and GONG observational networks \citep{2009ApJ...699.1403B}. This sound speed is consistent with a previous sound speed inversion computed from high accuracy data obtained by the GOLF and MDI instruments of the SOHO mission \citep{1997SoPh..175..247T}. 
In the solar core, the impact of $\eta-$parametrised dark matter on the sound speed is
much larger than can be accommodated by the current solar standard model \citep{2011RPPh...74h6901T}.
Moreover, although 
the sound speed is not inverted in the very central region of the Sun \citep{2009ApJ...699.1403B}, 
the reduction in the sun's speed profile caused by the presence of dark matter 
is much larger than previously tentative sound speed inversions of the Sun's core \citep{2011RPPh...74h6901T}.
As shown in figure 3, these dark matter models produce a decrease in the square of the sound speed 
of the order of  3\%-5\%, well above 1\% of the uncertainty of the current solar model.
This reduction in temperature is accompanied by significant changes in the Sun's core structure, 
which leads to  visible effects on other type of seismic diagnostics, such as the small acoustic mode separation\citep{1994A&A...290..845L,1993ApJ...408..347T}.
\citet{2010PhRvD..82j3503C} have found that dark matter particles that accumulate 
inside the  Sun's core and produces a temperature variation of this order of magnitude, 
also have a small acoustic mode separation quite different from helioseismology data.  
  
The relationship between the solar neutrino fluxes and the sound speed with the temperature is easily established,
if we notice that the $^8B$ and $^7Be$ neutrino fluxes depend on the central temperature $T_c$ as $T_c^{24}$ and $T_c^{10}$  \citep{1996PhRvD..53.4202B,2002PhRvC..65b5801B}, and the square of sound speed can be expressed as $C_s^2 \approx T_c/\mu_c $ where $\mu_c$ molecular weight  in the centre  \citep{1993ApJ...408..347T}. This different sensitivity to temperature explains why the diagnostics provided by  $\Phi(^8B)$, $\Phi(^7Be)$ and $C_s^2$ are 
quite distinct. We have also computed the solar neutrino fluxes of other pp-nuclear reactions and the results follow a similar behaviour. Other seismic diagnostic are possible, such as comparing the observed acoustic oscillations to the theoretical  acoustic oscillations, or their respective small separations. Nevertheless, 
the sensitivity of seismic quantities to changes in the Sun's core structure is always less visible than in solar neutrino fluxes.

\section{Discussion}

The current standard solar model
\citep{1993ApJ...408..347T,2010ApJ...715.1539T,2011ApJ...743...24S} is in agreement with the  $\Phi(^8B)$ and $\Phi(^7Be)$ neutrino fluxes measured by SNO and Borexino detector
\citep{2010PhRvC..81e5504A,2010PhRvD..82c3006B,2011PhRvL.107n1302B,2012PhRvL.108e1302B,Arpesella:2008vd}. 
The current $^8B$ neutrino flux observational determination in the case of 
no-neutrino oscillations (or electron neutrinos; including 
the theoretical uncertainty of neutrino flux solar neutrinos in the solar standard model) is 
$\Phi(^8B) =5.05_{-0.20}^{+0.19} \times 10^6\; {\rm cm^{-2} s^{-1}}$ for the SNO experiment \citep{2010PhRvC..81e5504A}
and  $\Phi(^8B) =5.88\pm 0.65 \times 10^6\; {\rm cm^{-2} s^{-1}}$ for the Borexino experiment 
\citep{2010PhRvD..82c3006B}. The Borexino experiment measures $^7Be$ solar neutrino flux 
$\Phi(^7Be) =4.87\pm 0.24  \times 10^9\; {\rm cm^{-2} s^{-1}}$, 
 under the assumption of the MSW-LMA scenario of solar neutrino oscillations \citep{2011PhRvL.107n1302B,Arpesella:2008vd}.

The $\Phi(^8B) $ neutrino measurements made by Borexino and SNO experiments suggest that the Sun's core is slightly hotter than expected, since the $\Phi(^8B) $ measured value is higher than the value predicted by the current solar standard model. This discrepancy is validated by the Borexino measurement of $\Phi(^7Be) $  which is sensitive to a region slightly off the Sun's centre. The first measurement of the pep neutrinos was done recently by the Borexino experiment, $\Phi(pep) =1.6\pm 0.3  \times 10^8\; {\rm cm^{-2} s^{-1}} $ \citep{2011PhRvL.107n1302B}. The experimental value also suggests that the Sun is hotter than the solar standard model predicts. 
This diagnostic is quite reliable once we realise that $\Phi(pep) $ is strongly dependent on the luminosity of the star. All these independent solar neutrino experiment results suggest that the solar standard model is cooler than the actual Sun.

The internal structure of the Sun is well known by means of solar neutrinos and helioseismology data. Therefore, the theoretical uncertainty of the standard solar model is consistently taken into account by comparing the predictions of these two probes with observations.
Several authors have shown that an important source of uncertainty on the calculation of the solar neutrino fluxes (electronic flavour) comes from the unclear measurements of heavy element abundances in the Sun's surface 
\citep{2005ApJ...621L..85B,2005ApJ...618.1049B,2009ARA&A..47..481A}. 
One other source of uncertainty is related with the determination of several pp-reaction rates and electron screening
\citep{2011ApJ...729...96M,2011ApJ...743...24S,2001A&A...371.1123W}. Similarly, other standard and no-standard physical processes that contribute
for the evolution of the star can also be an important source of uncertainty. Among others we can refer to the following indirect processes:  convective overshoot, low-z accretion, mass loss, solar rotation and meridional circulation \citep{2010ApJ...713.1108G,2010ApJ...715.1539T,2011RPPh...74h6901T}.  These suggestions have been made mostly to resolve the discrepancy between the theoretical sound speed as obtained 
with the new solar abundances  \citep{2009ARA&A..47..481A,Asplund:2005uv} and the sound speed obtained from the acoustic oscillations. 
As previously mentioned such physical processes have a much smaller effect on solar neutrino fluxes 
than the accumulation of dark matter in the Sun's core.

In the study of the impact of dark matter in the Sun, we will choose to consider an interval of theoretical uncertainty that takes into account such processes. Therefore, considering both the theoretical and experimental uncertainties, we choose to rule out models that predict a $^8B$ neutrino flux which deviates more than 30\% from our solar standard model. Similarly, a $^7Be$ neutrino flux of solar models which deviates more than 15\% from our solar standard model can also be ruled out. This is consistent with the fact that $\Phi(^8B) $  is two times more sensitive to the central temperature than $\Phi(^7Be) $ . The choice of this threshold is in agreement with other authors
\citep[e.g.,][]{2010PhRvD..82h3509T}. Furthermore, such intervals of uncertainties on $^8B$ and $^7Be$ also include the solar structure variations (e.g. temperature) in the Sun's interior, and in particular in the Sun's core that can be attributed to some of the physical processes previously mentioned.

Figure 2 shows the neutrino flux variations in  more than 30 evolution models of the Sun computed 
for different values of $\eta$  and $\langle \sigma v\rangle$.
This result is identical for both (s-wave and p-wave) annihilation channels.
In the $\eta $ vs. $\langle \sigma v\rangle$ plane, the iso-contours 
corresponding to $\Phi_\nu(^8B)$ and $\Phi_\nu(^7Be)$ show neutrino flux variations 
relatively to the solar standard model. 
We  choose to present  the case of  dark matter particles that have  $m_\chi=10$ GeV, 
spin-dependent scattering cross-section $\sigma_{SD}=10^{-40}\;{\rm cm^2}$,
and  spin-independent  scattering cross-section $\sigma_{SI}=10^{-36}\;{\rm cm^2}$. 
The neutrino flux variation shown follows the decrease of the local halo density of particles $\rho_\chi$, 
which follows the relic density $\Omega_\chi$ specific to each set of values ($\eta$, $\langle \sigma v\rangle$). 
 The observed difference in sensitivity between 
$\Phi_\nu(^8B)$ and $\Phi_\nu(^7Be)$ neutrino fluxes,  
is caused by the fact that  $^8B$ neutrinos are produced in a more central region than  $^7Be$ neutrinos. 

 If we adopt the 30\%  and 15\% fixed thresholds for the $^8B$ and $^7Be$ neutrino fluxes,  
 we conclude that we can rule out $\eta-$asymmetric dark matter particles with a mass of $10 \rm GeV$,
that have  an $\eta \ge 10^{-12}$ and an annihilation rate (both annihilation channels) larger than $10^{-23}\; {\rm cm^3 s ^{-1}}$. 
An analysis of other solar neutrino fluxes such as $pp$ neutrinos re-enforces the conclusions reached in this study.

\section{Conclusions}

In this paper, we proposed a new strategy to study the impact of dark matter in the evolution
of Sun and stars. We started by computing the basic cosmological primitive model that is responsible
by the formation of the dark matter asymmetric particles, assuming that dark matter like
baryons has an asymmetry, analogous to the baryon asymmetry. As a consequence the 
population of dark matter particles and antiparticles in the current Universe is fixed by the 
measured dark matter density $\Omega_{DM}h^2$. The amount of particles and antiparticles 
depends on the dark matter asymmetry parameter $\eta$, as well as on the dark matter 
annihilation (p- and s-wave) channels. Considering that the Sun is formed in a dark matter halo 
that has the same amount of particles and anti-particles, we computed the evolution of
the Sun in such conditions. 

Using the solar standard model as a reference model
for the internal structure of the Sun, we have studied the impact of asymmetric dark matter
on the production of solar neutrino fluxes. Following a procedure identical to the analysis 
proposed in \citet{2010Sci...330..462L,2010ApJ...722L..95L}, we
describe the impact of asymmetric dark matter on the Sun's core.

Several evolution models of the Sun were computed for a range of light dark matter particles. 
Particles with masses smaller than 5 GeV are not considered since evaporation becomes important in this mass range
and a large number of dark matter particles escape the gravitational field of the star,  significantly reducing the impact 
on the Sun's core \citep{1990ApJ...356..302G}. Similarly, particles with a mass above $20$ GeV produce
a very small dark matter core and their effect in the Sun's structure is almost negligible  \citep{2011PhRvD..83f3521L}.
Accordingly, dark matter particles with masses between 5 and 20 GeV,
produce a temperature decrease in the Sun's core that visibly affects the neutrino fluxes, 
although these depend on the values of specific parameters, such as the dark matter asymmetry  
and  annihilation cross section. 
The impact of more massive dark particles in the evolution of a star like the Sun should
be significantly smaller than in the case discussed in this work. The reason is the fact that the 
concentration of dark matter in the centre of the star, is usually characterized by a
dark matter core radius\citep{2009ApJ...705..135C}  which is inversely proportional to the square  
root of the mass of the dark matter particle. The choice of other parameters of the 
scattering cross section will not change fundamentally such results. 

We note incidentally that high annihilation rates for light dark matter particles are ruled out if accompanied by lepton production
from observations of Cosmic Microwave Background temperature fluctuations and of searches for gamma rays from nearby dwarfs. 
But these limits are irrelevant once asymmetric dark matter is important,  and most importantly for us, assume specific annihilation channels. Our solar constraints come from  accumulation of dark matter particles by the Sun, and complement other annihilation limits. The annihilations are of course  important for fixing the relic abundance. However a key point of our paper is that we consider asymmetric dark matter for which the dependence on specific annihilation channels is very weak. Furthermore,  the relic abundance is fixed by specifying the annihilation rate via freeze-out physics.

Indeed,  freeze-out may be far more complicated than given by the simple connection that specifies the "thermal" freeze-out 
cross-section in terms of the observed relic abundance. Many models violate this condition. Our models test independently 
of freeze-out the very important parameter space of annihilation rate today (in the low energy universe) whereas relic abundances 
probe the annihilation rate in a model-dependent way at freeze-out (at high energies). Moreover the Sun probes asymmetric dark 
matter at the present solar radius, and complements the use of dwarfs  in the halo and of Cosmic Microwave Background in the 
early universe for testing Majorana-type dark matter.

In this work we have computed the $^8B$ and $^7Be$  neutrino flux variations 
for several $\eta-$parametrised asymmetric dark matter particles,
corresponding to different values of $\eta$  and $\langle \sigma v\rangle$.
The values of  $\eta$  and $\langle \sigma v\rangle$  were chosen in such a way that $\Omega_{DM}$ for 
each of the asymmetric dark matter scenarios considered is consistent with current observational measurements 
(cf. Figure 1). We have presented the results of light dark matter  particles (typically 10 GeV),
with the same characteristics as the candidates ''observed'' by DAMA/LIBRA and CoGeNT experiments. 
We find that such type of particles produces large variations on the flux of solar neutrinos.
The $^8B$ neutrino flux variation takes values between 35\% and 65\% for s-wave and p-wave annihilation channels, 
clearly above the uncertainty in the current solar models estimated to be of the order of 15\%  \citep{2011RPPh...74h6901T}. 
This conclusion holds even for our more conservative threshold of 30\% on $^8B$ neutrino flux
or a threshold of 15\% on $^7Be$ neutrino flux, as discussed in the previous section.
Likewise, dark matter particles with $m_\chi=15$ GeV show $^8B$ neutrino flux variations of the order of $16\%$ up to $ 40\%$,
which can be rejected if we consider the 15\% threshold of uncertainty in the physics of the solar standard model. Therefore, it is reasonable to consider that light dark matter particles that have a scattering cross section of the order of a pico barn, as suggested  by several theoretical models to explain the DAMA/LIBRA and CoGeNT experiments, produce neutrino fluxes that are in disagreement 
with the current neutrino flux observations. On this basis, assuming that stellar evolution is affected by dark matter as discussed here, 
it is reasonable to assume that such
types of particles cannot exist in the current Universe. Forthcoming solar neutrino flux experiments could restrain even further the
parameters of the $\eta-$asymmetric dark matter candidates proposed by the recent dark matter models.
 
\begin{acknowledgments}
This work was supported by grants from "Funda\c c\~ao para a Ci\^encia e Tecnologia"  and "Funda\c c\~ao Calouste Gulbenkian" and the NSF (grant OIA-1124453). We would like to thank the referee for the useful comments and insightful suggestions that improved the quality of the paper. 
\end{acknowledgments}
%

\end{document}